\documentclass{sig-alternate-10pt}
\usepackage[font=bf]{caption}

\newcommand{\subparagraph}{}

\usepackage[compact]{titlesec}

\usepackage{multirow}
\usepackage[sort,nocompress]{cite}
\usepackage[usenames,dvipsnames]{color}
\usepackage{amsfonts}
\usepackage{times}
\usepackage{xspace} \usepackage{epsfig}
\usepackage{amsmath}
\usepackage{amssymb}
\usepackage[hyphens]{url}
\usepackage{listings}
\usepackage{fancyvrb}
\usepackage{graphicx}
\usepackage{subfig}
\usepackage[english]{babel}
\usepackage{courier}

\usepackage{listings}
\VerbatimFootnotes

\newcommand{\tightcaption}[1]{\vspace{-0.2cm}\caption{#1}\vspace{-0.4cm}}

\newcommand{\comment}[1]{}
\newcommand{\mycomment}[1]{}
\newcounter{note}[section]

\newcommand{\Section}{\S}

\usepackage{pifont}

\newcommand{\mypara}[1]{\smallskip\noindent{\bf {#1}:}~}
\newcommand{\myparatight}[1]{\smallskip\noindent{\bf {#1}:}~}

\newcounter{packednmbr}

\newenvironment{packedenumerate}{\begin{list}{\thepackednmbr.}{\usecounter{packednmbr}\setlength{\itemsep}{0.5pt}\addtolength{\labelwidth}{-4pt}\setlength{\leftmargin}{\labelwidth}\setlength{\listparindent}{\parindent}\setlength{\parsep}{1pt}\setlength{\topsep}{0pt}}}{\end{list}}

\newenvironment{packeditemize}{\begin{list}{$\bullet$}{\setlength{\itemsep}{0.5pt}\addtolength{\labelwidth}{-4pt}\setlength{\leftmargin}{\labelwidth}\setlength{\listparindent}{\parindent}\setlength{\parsep}{1pt}\setlength{\topsep}{0pt}}}{\end{list}}

\begin{document}

\title{Analyzing TCP Throughput Stability and Predictability with Implications for Adaptive Video Streaming}
\author{Yi Sun$^\circ$, Xiaoqi Yin$^\dag$, Nanshu Wang$^\circ$, Junchen Jiang$^\dag$, Vyas Sekar$^\dag$, Yun Jin$^\diamond$, Bruno Sinopoli$^\dag$\\ $^\circ$ICT, CAS $^\dag$ Carnegie Mellon University $^\diamond$ PPTV}

\maketitle


\begin{abstract}
\label{sec:abstract}

Recent work suggests that TCP throughput stability and predictability within a
video viewing session can inform the design of better video bitrate adaptation
algorithms.  Despite a rich tradition of Internet measurement, however, our
understanding of throughput stability and predictability is quite limited.  To
bridge this gap, we present a measurement study of throughput stability  using
a large-scale dataset from a video service provider.  Drawing on this analysis,
we propose a  simple-but-effective  prediction mechanism based on a hidden
Markov model and demonstrate that it outperforms other approaches.  We also
show the  practical implications in improving the user experience of adaptive
video streaming.

\end{abstract}
\section{Introduction}
\label{sec:introduction}

In recent years, we have seen a  dramatic rise in the volume of HTTP-based
adaptive video streaming  traffic in the Internet~\cite{ciscovni}. In contrast
to traditional  metrics such as transfer completion time for web requests,
delivering good application-level experience for video introduces new metrics
such as low buffering or smooth bitrate
delivery~\cite{dobrian2011understanding}.

 To meet these new application-level quality of experience goals, video players
use dynamic bitrate adaptation within a viewing
session~\cite{huang2014dash,jiang2014improving}.  Here, the video is chunked
into discrete segments, and each chunk is encoded at different bitrate levels,
to enable the player to dynamically change the bitrate chosen for future video
chunks in response to the operating conditions~\cite{huang2014dash,jiang2014improving}.  Note that in this setting,
delivering good application performance depends on the ``consistency'' of TCP
throughput behavior within the session between the client and the video server,
rather than the burst or average properties of the Internet path.

 In this respect, understanding intra-session TCP throughput characteristics
 can improve our understanding of existing video adaptation strategies
(e.g.,~\cite{yin2015controlvideo}) and  inform the development of new
algorithms (e.g.,~\cite{yin2015controlvideo,hotmobile}).  Specifically, there
are two key questions that we wish to address:

\begin{packeditemize}

\item {\bf Stability:} If the TCP throughput is  stable, then adaptive video
streaming algorithms can avoid frequent switches and pick the highest  possible
 bitrate that does not induce buffering~\cite{yin2015controlvideo,jiang2014improving,confused}.

\item {\bf Predictability:} Many adaptation algorithms use the estimated  TCP
throughput from previous chunks to choose the bitrates for the next few chunks.
Recent work has shown that an accurate throughput predictor, if available, can
significantly improve the quality of experience for adaptive video
streaming~\cite{yin2015controlvideo, hotmobile}.

\end{packeditemize}

 Despite the rich measurement literature in characterizing various Internet
path properties
(e.g.,~\cite{jain2003end,bismark,dischinger2010glasnost,hu2004pathneck}), our
understanding of TCP throughput stability and predictability is quite limited.
There has been surprisingly little work in this space and the closest related
works we are aware of are dated and limited in
scope~\cite{zhang2001constancy,wideareastability}.

Our goal in this paper is to bridge this gap. To this end,
  this paper makes three key contributions:
\begin{packeditemize}

\item {\em Measurement (\Section\ref{sec:analysis}):} We analyze the TCP
throughput stability on a dataset consisting of minute-level throughput
measurements from over 200K sessions from a large video provider. Our
key findings are: a) A large number of sessions have significant intra-session
throughput variations; b) High throughput sessions tend to be more stable; and
c) The throughput is more similar in neighboring/recent time slots and less
similar to measurements made further apart.

\item  {\em Prediction algorithm (\Section\ref{sec:premodel}):} Building on
observed temporal structure,  we develop a simple-yet-effective algorithm based
on the insight  
that the throughput can be modeled as a function of a hidden
state variable --  the number of concurrent flows at a bottleneck link. We
develop a hidden Markov model (HMM) predictor and show that it outperforms a range of
 timeseries modeling techniques.

\item {\em Application implications (\Section\ref{sec:eval}):}  Using
trace-driven simulations, we  show  that our HMM predictor significantly
improves the video QoE over prior work that does not use throughput
predictions~\cite{huang2014dash} and is very close to the optimal achievable
QoE which is based on the perfect knowledge of future throughput.

\end{packeditemize}

\section{Related work}
\label{sec:relwork}
 In this section, we place our work in the context of past work in Internet
measurement and adaptive video streaming.

\myparatight{Measuring path properties}  Prior work has  measured stability of path properties
such as  the persistence and prevalence of routes over
time~\cite{paxson1996sigcomm}.  Other work focuses on inter-domain routing
stability and reports that popular destinations have
stable routes~\cite{rexford2002bgp}.  In contrast our focus is on throughput
stability and predictability.

\myparatight{Bandwidth measurement tools} There are many tools for measuring the
available bandwidth and the capacity of Internet paths
(e.g.,~\cite{pathchar,hu2004pathneck}).    At a
high-level, they  extend packet pair techniques and provide mechanisms  to deal
with background traffic interference.  We refer readers
to the survey  by Jain and Dovrolis for more in-depth comparisons~\cite{jain2003end}. However, these are active probes that
result in a single data point. In contrast, we use passive measurements to
develop a systematic understanding of the temporal stability and predictability
of the TCP throughput.

\myparatight{Throughput stability} Balakrishnan et al., use throughput measurements
from a large web service and report that the throughput for the same
client-server pair does not change significantly (less than factor of 2) for
tens of  minutes~\cite{wideareastability}. Zhang et~al., analyze the stability
in terms of statistical, operational, and predictive
metrics~\cite{zhang2001constancy}. They report that using recent history on
the scale of minutes is useful but in the order of hours misleads estimators.
Unfortunately, these are dated and limited in terms of
scale and scope.\footnote{These were
performed in the late 90s and early 2000s and pre-date the widespread
deployment of high-speed broadband, CDNs, and the growth of Internet video.
Furthermore, they focus on a handful of source-destination pairs mostly located in
universities.}

\myparatight{Throughput prediction} Prior work developed approximate analytical models of
TCP throughput  as a function of  packet loss and delay~\cite{padhye1998modeling,tcppredictability,mirza2007machine}. However, these do not directly translate into actual prediction algorithms  that can feed into video adaptation algorithms.

\myparatight{Broadband measurements} Given the recent debate on network neutrality
and video, measurements of broadband  characteristics have regained
prominence~\cite{bismark,fccmba}. However, these do not focus on throughput
stability and predictability.

\myparatight{Adaptive video streaming over HTTP} Our work is motivated by Dynamic
Adaptive Streaming over HTTP (DASH). Prior work implicitly assumes throughput
is unstable and unpredictable and eschews this in favor of using the player
buffer occupancy for controlling bitrates~\cite{huang2014dash}. Recent work
~\cite{yin2015controlvideo,hotmobile} argues that adaptive video streaming can
significantly benefit from accurate throughput prediction. However, these do not
 provide  a concrete prediction algorithm. Our
contribution is in developing an effective throughput predictor and
demonstrating its utility for DASH.

\section{Dataset}
\label{sec:dataset}

In this section, we describe the dataset  we use for analyzing TCP throughput
stability and predictability.

Note that in contrast to other throughput and path measurements,  we need
continuous measurements over sufficiently long durations (e.g., several
minutes).  We are not aware of  public datasets that enable such in-depth
analysis of throughput stability and predictability at scale. We explored
datasets such as Glasnost~\cite{dischinger2010glasnost}, FCC~\cite{fccmba}, and
from a EU cellular provider~\cite{hsdpa}. Unfortunately, all of these
had too few hosts and  the sessions lasted only a handful of seconds making it
unusable for the stability and predictability analysis of interest for adaptive
video streaming.

\begin{figure}[t]
\vspace{-0.4cm}
\centering
\subfloat[Duration]
{
\includegraphics[width=0.5\linewidth]{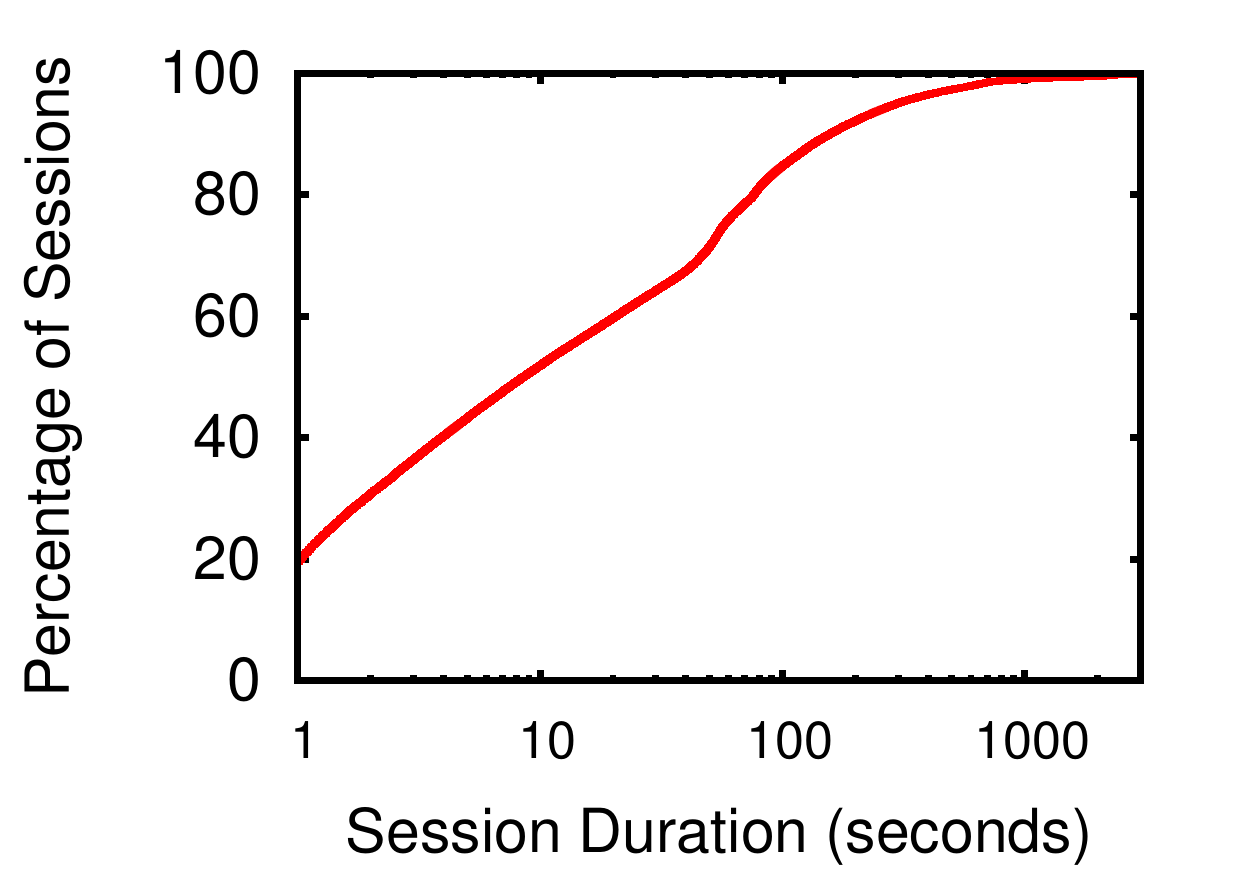}
\label{fig:duration}
}
\subfloat[Throughput]
{
\includegraphics[width=0.5\linewidth]{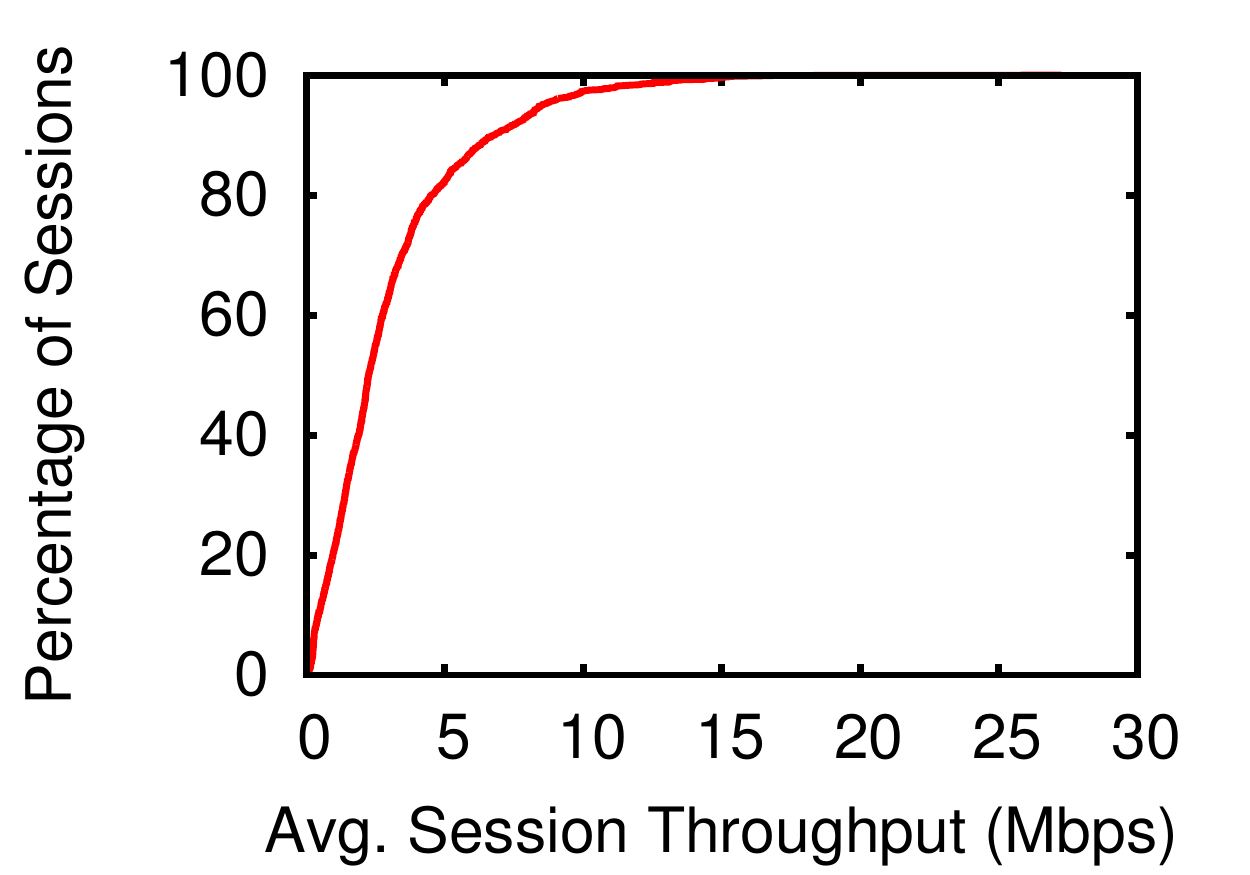}
\label{fig:speed}
}
\tightcaption{CDF of session duration and throughput.}
\label{fig:session}
\end{figure}

 Our dataset is collected from the operational CDN platform of
PPTV~\cite{pptv}. PPTV is a leading online video content provider in China with
more than 227 million users. We use measurements from real video sessions from
this provider.  These sessions cover $428,000$ unique client IPs  and over
1000 unique server IPs.  The clients span 508 cities and 28 ISPs in China.  In
total, we collect data from  over 2.7 million sessions over a 4 day period.

 Each session consists of several video ``chunks''. The session is divided into
1-minute epochs, and  the client reports the   average TCP throughput observed
within this epoch during the active download times. If there are  multiple
chunks transmitted in the same epoch, the throughput reported for this epoch
should be the byte-weighted mean of the average throughput of each chunk.
Conversely, if a chunk spans multiple epochs it contributes partially to each
epoch it spans.

 As observed in other studies, the duration of each session is
variable~\cite{videoqoehotnets}.  Figure~\ref{fig:duration}  shows the CDF of
the session duration in our dataset.  Since we are interested in temporal
stability and predictability, we focus on sessions that last more than 6
minutes.  About 10\% of the sessions last more than 6 minutes still yielding a
substantial number of sessions ($\geq$ 200K) for our analysis.
Figure~\ref{fig:speed} shows the CDF of the per-epoch average throughput and
suggests that the average throughput distribution is similar to residential
broadband characteristics~\cite{bismark}. While this is indeed a single dataset
from the Chinese Internet, based on these observations and our experience with
other datasets of a similar nature (e.g.,~\cite{dobrian2011understanding}) we
believe that this is representative of video workloads measured in
residential broadband settings.

 We do acknowledge one limitation---the finest time resolution we have is  1
minute. However, we believe that understanding stability/predictability at a
minute timescale is still valuable for adaptive video streaming applications
and as we will show in \Section\ref{sec:eval} it can still yield significant improvements for quality of
experience.

\begin{figure*}[thbf]
\vspace{-0.5cm}
\subfloat[Absolute stddev]
{
\includegraphics[width=0.32\linewidth]{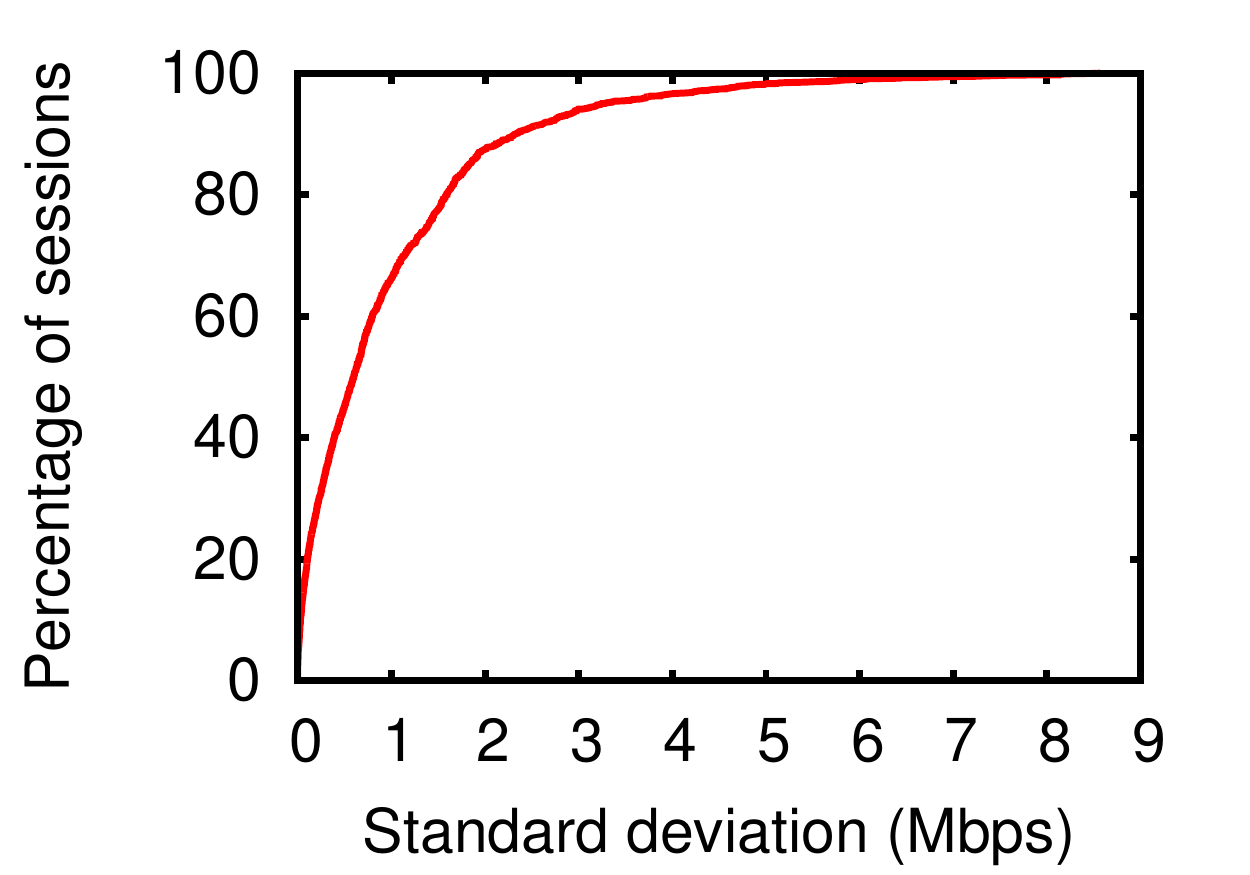}
\vspace{-0.2cm}
\label{fig:stddev}
}
\subfloat[Coefficient of variability]
{
\includegraphics[width=0.32\linewidth]{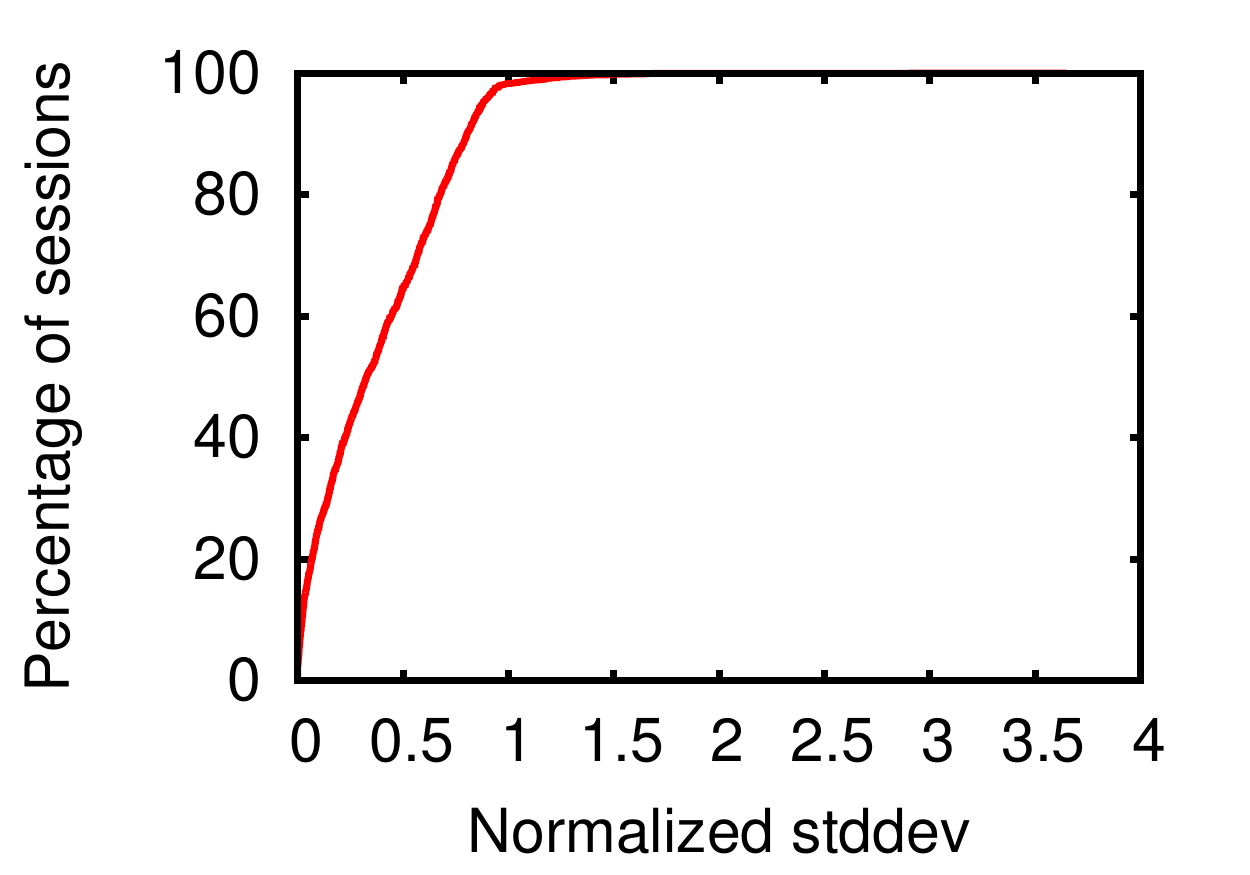}
\vspace{-0.2cm}
\label{fig:nstddev}
}
\subfloat[Diff.\ between 75-th and 25-th percentile]
{
\includegraphics[width=0.32\linewidth]{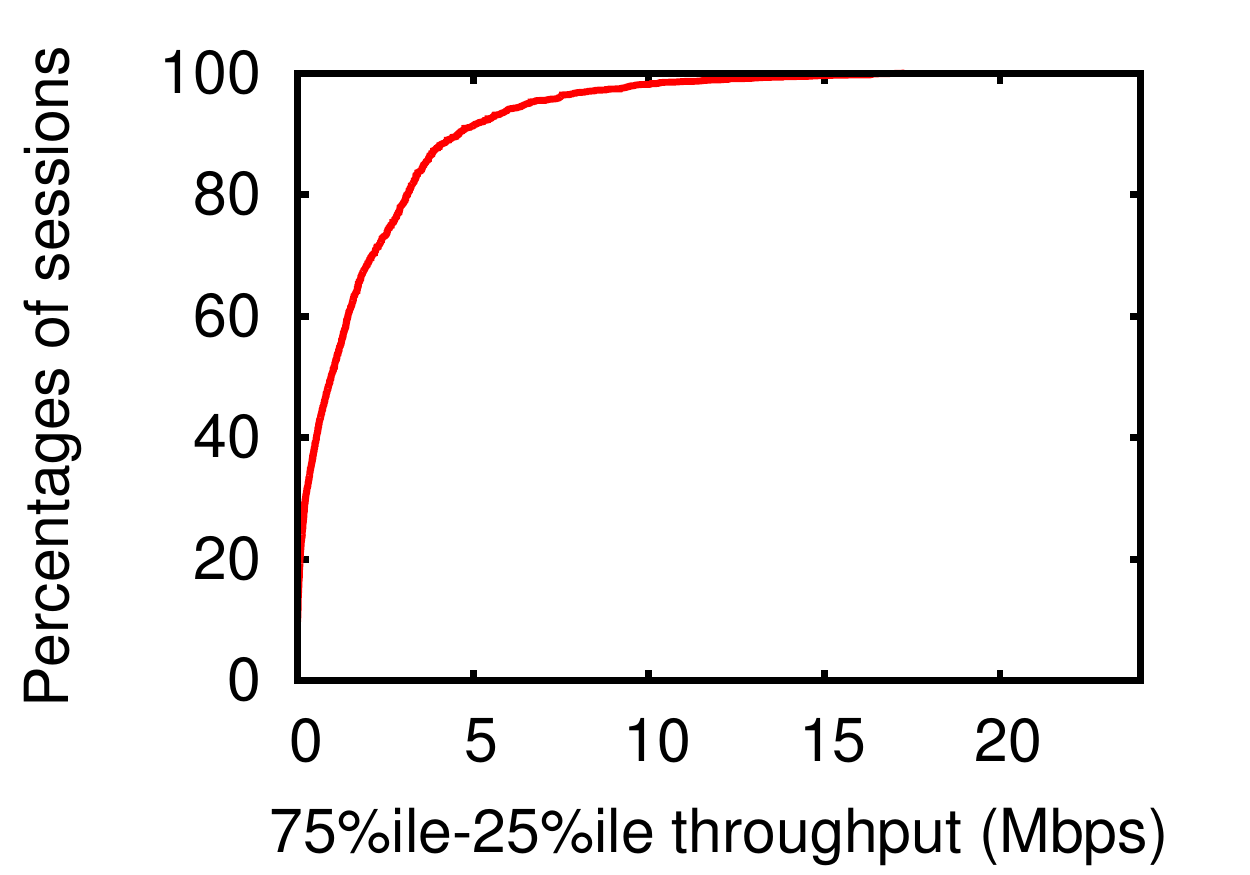}
\vspace{-0.2cm}
\label{fig:difference}
}
\tightcaption{Analyzing throughput stability through different metrics.}
\end{figure*}

\section{Intra-session throughput analysis}
\label{sec:analysis}

 In this section, we analyze three key characteristics of the
 throughput within  a client-server session:

\begin{packedenumerate}

\item How {\em variable} is the throughput within a session? \\ For instance,
if the variability is small, then the adaptation logic does not have to switch bitrates often.

\item Is the {\em variability} correlated/anti-correlated vs.\  average
throughput? \\ If the variability is a function of the average throughput, then
we may need to customize the adaptation logic for different deployments; e.g.,
wireless clients  vs.\  fiber-to-home links.

\item Are there {\em temporal patterns} within the session; e.g., how similar
are recent observations made $k$ minutes apart? \\ This temporal structure has
key implications for predictability as many adaptation algorithms use
estimates of throughput over the next few chunks as part of their decision
logic~\cite{yin2015controlvideo}.

\end{packedenumerate}

\mypara{Intra-session variability} First, we compute the standard deviation
(``stddev") of TCP throughput across  different measurements within the
session.  Figure~\ref{fig:stddev} shows the CDF (across sessions) of the
per-session throughput stddev.  We see that about 20\% of sessions have a
stddev $\ge$2Mbps.  Second, we compute the coefficient of variation, which is
the ratio of stddev to the mean.  Figure~\ref{fig:nstddev} shows the CDF of
this normalized metric; we see that the roughly 40\% of sessions have
normalized stddev $\ge$50\%.  Now, the stddev could still be biased by a few
outliers even if the throughput is mostly stable.\footnote{For instance,
consider a session with measurements 2,2,2,2,20. This will have a very high
stddev even though it is mostly stable.} Thus, we also compute the difference
between the 75-th and 25-th percentile throughput values within a session and
plot the CDF in Figure~\ref{fig:difference}. Again, we see that a non-trivial
fraction of sessions ($\geq$ 30\%) has a difference of $>$2Mbps/s.  In short,
this result confirms the general perception that we need good bitrate
adaptation strategies and that simple
static bitrate selection will not suffice.

\begin{figure}[h]
\vspace{-0.5cm}
\centering
\includegraphics[width=0.8\linewidth]{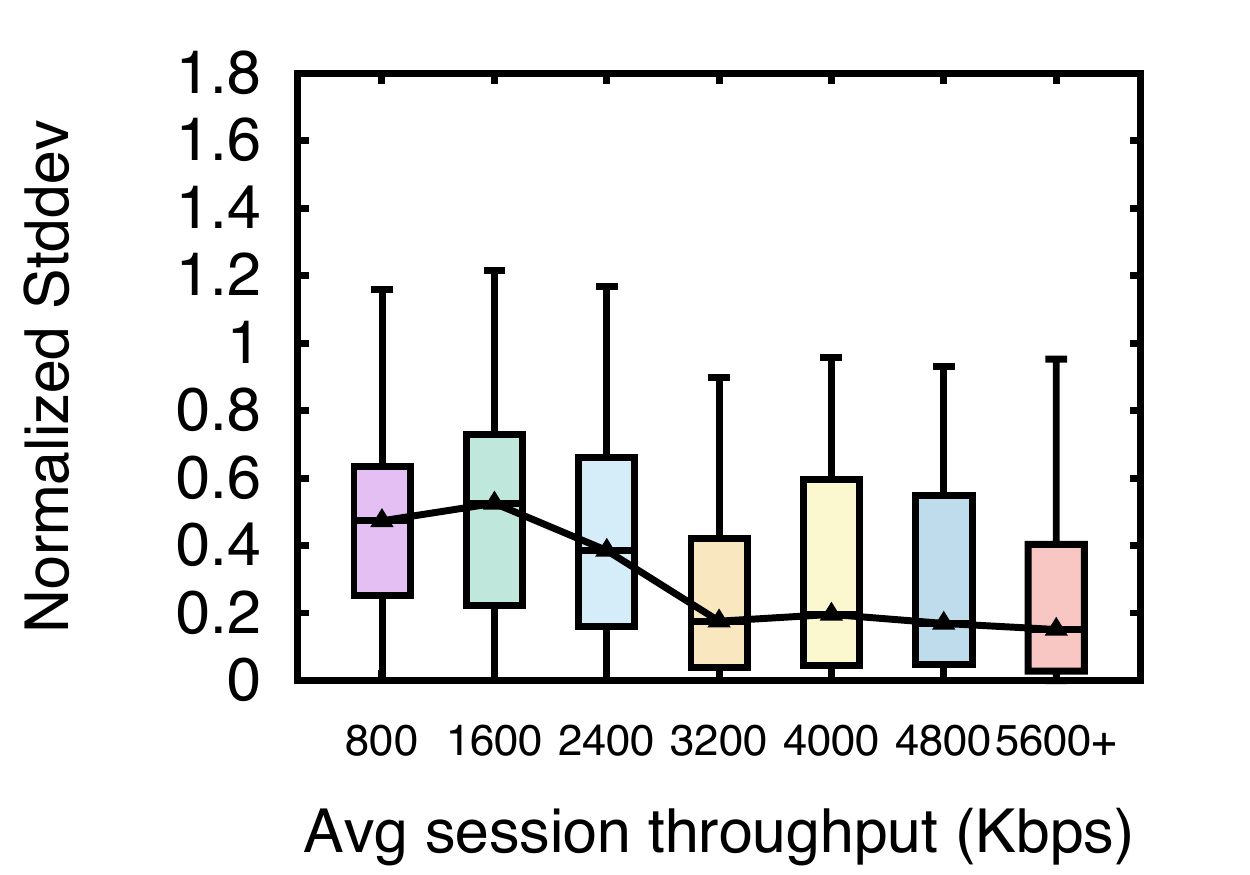}
\vspace{-0.1cm}
\tightcaption{Normalized stddev vs.\ average throughput.}
\label{fig:correlation:normalized}
\end{figure}

\mypara{Variability vs.\ average throughput}  Next, we analyze if there is some
relationship between throughput stability and the average session throughput.
Based on the  distribution in Figure~\ref{fig:speed}, we categorize the PPTV
sessions into different 800Kbps  bins.  Figure~\ref{fig:correlation:normalized}
shows the  average normalized stddev of the sessions within each bin.  As a
general trend, the normalized variability  decreases as with increased  average
throughput. We posit that  such high throughput  sessions traverse less
congested paths and thus the variability of throughput is also small.  This
result suggests that the throughput is  more stable for   higher
throughput sessions and thus bitrate adaptation algorithms can afford to be
less conservative compared to low throughput sessions.

\mypara{Temporal structure} The above results provide an aggregate view of the
variability within the session but do not shed light on the temporal structure.
 Such temporal  structure can have key implications for predictability.  For instance,
consider two hypothetical sessions with the following measurements (in Mbps):
(1) Session 1 = 1,1,1,0.5,0.5,0.5 and (2) Session 2 = 1,0.5,1,0.5,1,0.5.  Now,
both sessions have the same mean, stddev, percentile difference, but
intuitively Session 1 is more predictable based on recent history
 than the pattern  in Session 2.

\begin{figure}[h]
\vspace{-0.5cm}
\centering
\includegraphics[width=0.8\linewidth]{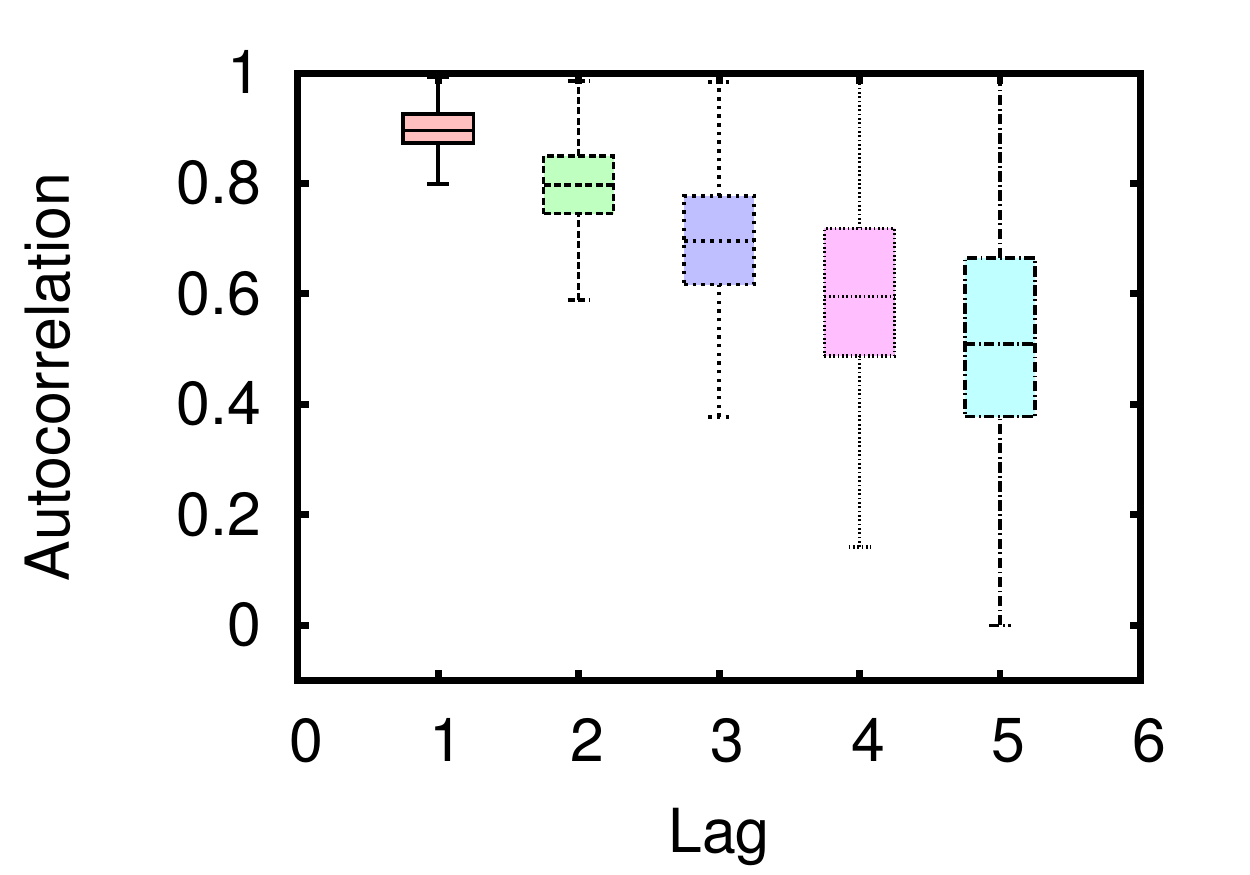}
\vspace{-0.1cm}
\tightcaption{Autocorrelation of the throughput over different time windows (25, 50, 75\%ile).}
\label{fig:autocorrelation}
\end{figure}

To quantitatively analyze the temporal structure (i.e., how the throughput
changes during the course of a session),  we compute  the autocorrelation of
the throughput time-series for different time-shifts.\footnote{The
autocorrelation is defined as the $\mathit{R(\tau)} =
\frac{\mathbb{E}[(X_t-\mu)(X_{t+\tau}-\mu)]}{\sigma^2}$, where $X_t$ is the
throughput at time slot $t$, $\mu$ is the mean value of the throughput for the
whole session, and $\tau$ is the time lag in the time series.}
Figure~\ref{fig:autocorrelation} summarizes the distribution (across sessions)
of the these autocorrelations  as a box-and-whiskers plot depicting the median,
25-th, 75-th percentiles and the min/max values for different time lags.  While
the autocorrelations are positive, we see a marked decrease as the lag
increases. In other words, the throughput is more similar in recent
time slots and less similar to measurements made far earlier or later.

\begin{figure}[h]
\centering
\includegraphics[width=0.7\linewidth]{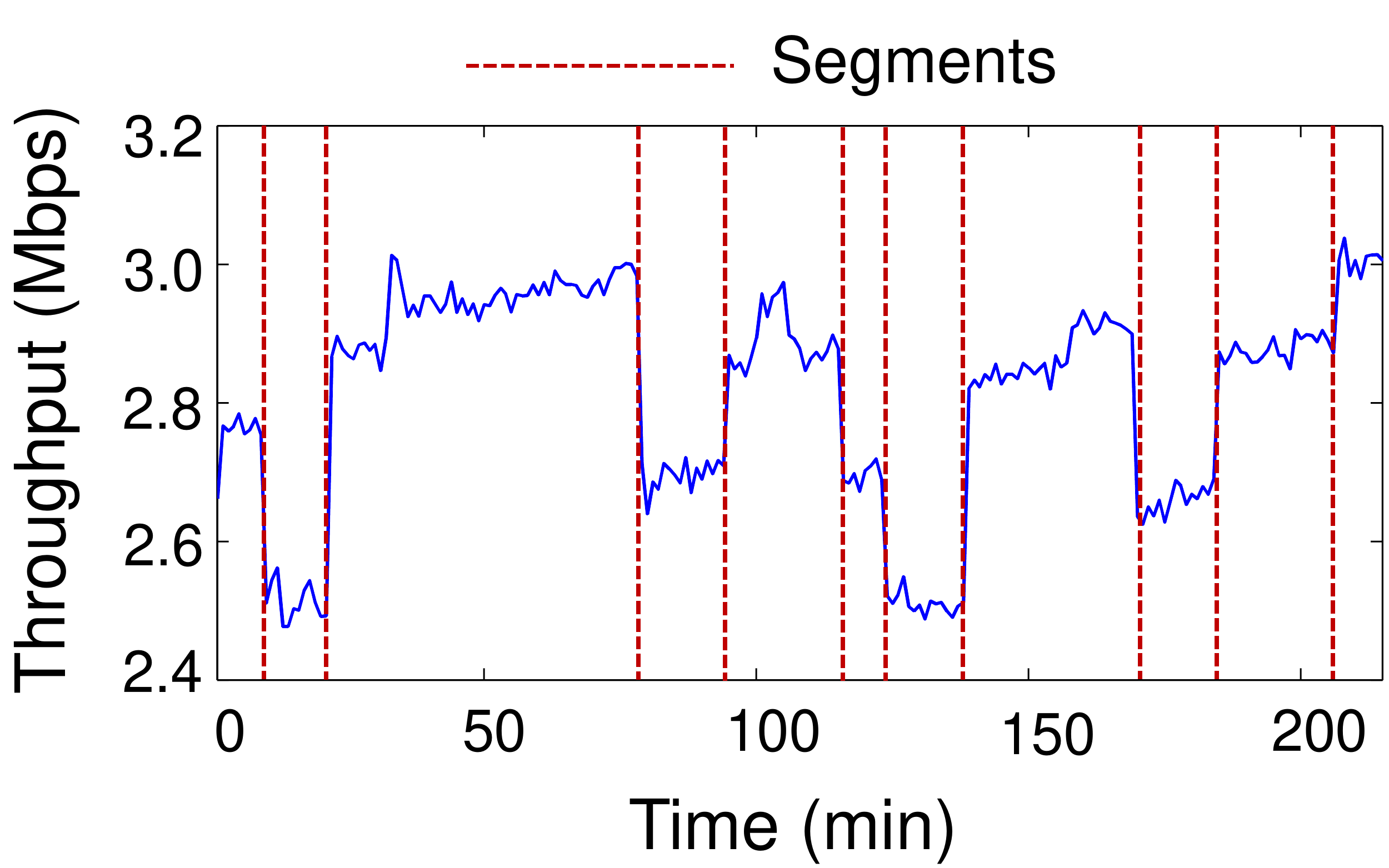}
\tightcaption{An example of PPTV video session throughput.}
\label{fig:example_session}
\end{figure}

To give some visual intuition, we show the throughput timeseries of a
representative client-server session in Figure~\ref{fig:example_session}.
Here, we see that the throughput evolves during the course of a session and
thus the correlation between distant timeslots tends to be lower.

\mypara{Summary of key findings:}
 Our throughput variability analysis shows that:
\begin{packedenumerate}
\item A large number of sessions have significant variations of their intra-session throughput, with  normalized stddev $\ge$50\%
 for more than 40\% of sessions.
\item High throughput sessions generally are more stable than low throughput sessions.
\item The throughput is more similar in recent measurements and the similarity decays
  with higher lag.
\end{packedenumerate}

\section{Intra-session throughput prediction}
\label{sec:premodel}

The previous section reveals  significant throughput variation during a session
and the need for  good video bitrate adaptation schemes.  Ideally, we can
accurately predict the TCP throughput to select bitrates for the next few
chunks to optimize user perceived quality of
experience~\cite{yin2015controlvideo, tian2012towards}. However, this is
challenging and the limitations of existing prediction mechanisms
(\Section\ref{sec:premodel:strawman}) have even motivated efforts that avoid
throughput-based adaptation~\cite{huang2014dash}.  In this section, we describe
a simple but effective prediction motivated by the temporal structure in the
throughput. Before we do so, we describe strawman solutions considered in the
literature and their limitations in light of our observations.

\subsection{Strawman solutions}
\label{sec:premodel:strawman}

Our goal here is not to exhaustively enumerate all possible prediction
algorithms. As such, the models  we consider are representative of classical
time series models used in adaptive streaming
proposals~\cite{yin2015controlvideo, tcppredictability}.\footnote{We also
 tried  ``forecast'' models that extrapolated trends but these
performed worse and are not shown.}  At a high level, a
throughput prediction model can be viewed as a function of the observed
throughputs over the previous  $p$ epochs.  Let $W_t$ denote the  observed
throughput at epoch $t$ and $\hat{W}_t,\ldots,\hat{W}_{t+\Delta}$ denote the
estimate for the next $\Delta$ epochs.

\begin{packeditemize}

\item \textbf{{Last Sample} (LS)}: In the simplest case, we simply use the previous
observation; i.e., $\forall i \in [t,t+\Delta]:  \hat{W}_i=W_{t-1}$. The
problem with this approach is that a single sample will be a very noisy
estimator and thus may cause significant bitrate oscillations~\cite{yin2015controlvideo,jiang2014improving}.

\item \textbf{Arithmetic Mean (AM)}: To address the noise, we can consider
``smoothing'' using  $p$ measurements from  history; i.e.,   $ \forall i \in
[t,t+\Delta]: \hat{W}_i=\frac{\sum_{q=1}^p W_{t-q}}{p}$.  However, there are
still two fundamental problems.  First, if we use a small $p$,  outliers can
still cause significant under- or overestimation. Second, if we use a large
$p$,  measurements made too far back in history may induce serious biases as we saw in
Figure~\ref{fig:example_session}.

\item \textbf{Harmonic Mean (HM)}: One way to minimize the impact of outliers
in AM is using a harmonic mean~\cite{jiang2014improving}: $\forall i \in
[t,t+\Delta]:\hat{W}_i = \frac{1}{\sum_{q=1}^p \frac{1}{W_{t-q}}}$. While this
addresses the outlier problem, uncorrelated  measurements
too far in  history can still bias the predictions.

\item \textbf{Auto-regressive models (ARMA,AR):} Auto-regressive moving average
(ARMA)  is a classical timeseries modeling technique~\cite{tcppredictability}.
The ARMA model assumes  $W_t$ has the following form: $ W_{t} = a_0 +
\sum_{j=1}^p a_j W_{t-j} + \sum_{j=1}^q b_j S_{t-j}$, where $S_t \sim
N(0,\sigma^2)$ is i.i.d. Gaussian noise,  independent of $W_t$.  $p,q$ are the
sizes of the sliding windows for auto-regression and moving average,
respectively, and $\theta_{ARMA} = \{\{a_i\}_{i=0}^p, \{b_i\}_{i=1}^q\}$ are
the parameters that can be learned from training data (e.g., historical
sessions).  The auto-regression (AR) model is  a simplified version of ARMA
that  assumes  $W_{t} = a_0 + \sum_{j=1}^p a_j W_{t-j} + e_t$, where $a_0$ is a
constant and $e_t$ is 
i.i.d. zero-mean Gaussian noise independent of $W_t$.
$\theta_{AR} = \{\{a_i\}_{i=0}^p\}$.  Given training data and
$p,q$, Yule-Walker equations can be adopted to learn the parameters
$\theta_{ARMA}$, or $\theta_{AR}$.  The key problem with these models is that
they have implicit independence and stationary assumptions.   However,
Figures~\ref{fig:autocorrelation} and~\ref{fig:example_session} suggest that
there is some inherent ``stateful'' and ``evolving'' temporal structure in the
throughput, which contradicts these assumptions.

\end{packeditemize}

\subsection{Using a Hidden Markov Model}
\label{sec:premodel:hmm}

Hidden Markov models (HMM) are widely used in many applications, ranging from
speech recognition to event detection \cite{bishop2006pattern}. From a
networking perspective, the intuition behind the use of HMM in our context is
that the throughput depends on the hidden state---the number of flows
sharing the bottleneck link.  The visualization in
Figure~\ref{fig:example_session} confirms this intuition that the throughput
has some stateful evolving behaviors.  By capturing these state transitions and
the dependency between the throughput vs.\ the hidden state, using HMM can
yield more robust throughput predictions.

\mypara{Model specification} Suppose the throughput depends on some hidden
state variables $X_t\in{\cal X}$, where ${\cal X} = \{x_1,\cdots, x_M\}$ is the
set of possible states and $M = |{\cal X}|$ is the number of states.  The state
evolves as a Markov process where the likelihood of the current state only
depends on the last state, i.e., $\mathbb{P}(X_t|X_{t-1},X_{t-2},\cdots, X_1) =
\mathbb{P}(X_t|X_{t-1})$. We denote the transition probability matrix by
$P=\{P_{ij}\}$, where $P_{ij} = \mathbb{P}(X_t = x_i|X_{t-1}=x_j)$.  We let the
probability distribution vector $\pi_t = (\mathbb{P}(X_t = x_1), \cdots,
\mathbb{P}(X_t = x_M))$.  Then $\pi_{t+\tau} = \pi_{t}P^{\tau}$.  Each state ``emits''
the throughput expected within that state. Within each  hidden state
$X_t$, we model  the  throughput $W_t$ by a  Gaussian distribution; i.e., $W_t|X_t=x \
\sim N(\mu_x, \sigma_x^2)$.

To see this concretely, let us revisit Figure~\ref{fig:example_session}.  Here,
we can conceptually think of splitting the timeseries into roughly 11
 segments each corresponding to a hidden  state.  Within each segment, the
throughput is largely Gaussian;  e.g., between timeslots 20--75 the throughput
has mean 2900, and in slots 10-20 and 125--135 the mean is  2500.

\mypara{Model learning} Given number of states $M$, we can  use training data
to  learn the parameters of HMM, $\theta_{HMM} = \{\pi_0, P, \{(\mu_x,
\sigma_x^2), x\in{\cal X}\}\}$ via the expectation-maximization (EM) algorithm
\cite{bishop2006pattern}.  Note that the number of states $M$ needs to be
specified. There is a tradeoff here in choosing suitable $M$.  Smaller $M$
yields simpler models, but may be inadequate to represent the space of possible
behaviors.  On the other hand, a large $M$ leads to more complex model with
more parameters, but may in turn lead to overfitting issues.   We find
empirically that $M=6$ is a ``sweet spot'' in the tradeoff (Figure~\ref{fig:hmmstate}).

\mypara{Online throughput prediction} At time $t$, given past throughput $W_{1:t-1}=\{W_1,\cdots,W_{t-1} \}$, we first use
forward-backward algorithm \cite{bishop2006pattern} to determine
$\pi_{t-1|1:t-1} = (\mathbb{P}(X_{t-1} = x_1|W_{1:t-1}), \cdots,
\mathbb{P}(X_{t-1} = x_M|W_{1:t-1}))$. Then the distribution of $X_{t+\tau}$ can be
obtained by: $\pi_{t+\tau|1:t-1} = \pi_{t-1|1:t-1}P^{\tau+1}$. Finally, we compute
 the maximum likelihood estimate of $W_{t+\tau}, \tau > 0$  as
$\hat{W}_{t+\tau} = \mu_x$, where $x = arg\max_{x\in{\cal X}}
\mathbb{P}(X_{t+\tau}=x|W_{1:t-1})$.

\section{Evaluation}
\label{sec:eval}

In this section, we present  trace-driven evaluations using the dataset in
\Section\ref{sec:dataset} and evaluate our proposed HMM scheme vs.\
strawman approaches along two  dimensions: (1) Prediction accuracy and (2)
Video quality of experience.

\subsection{Improvement in prediction accuracy}
\vspace{-0.2cm}
\label{sec:premodel:eval}

\begin{figure}[t]
\vspace{-0.5cm}
\centering
\includegraphics[width=0.6\linewidth]{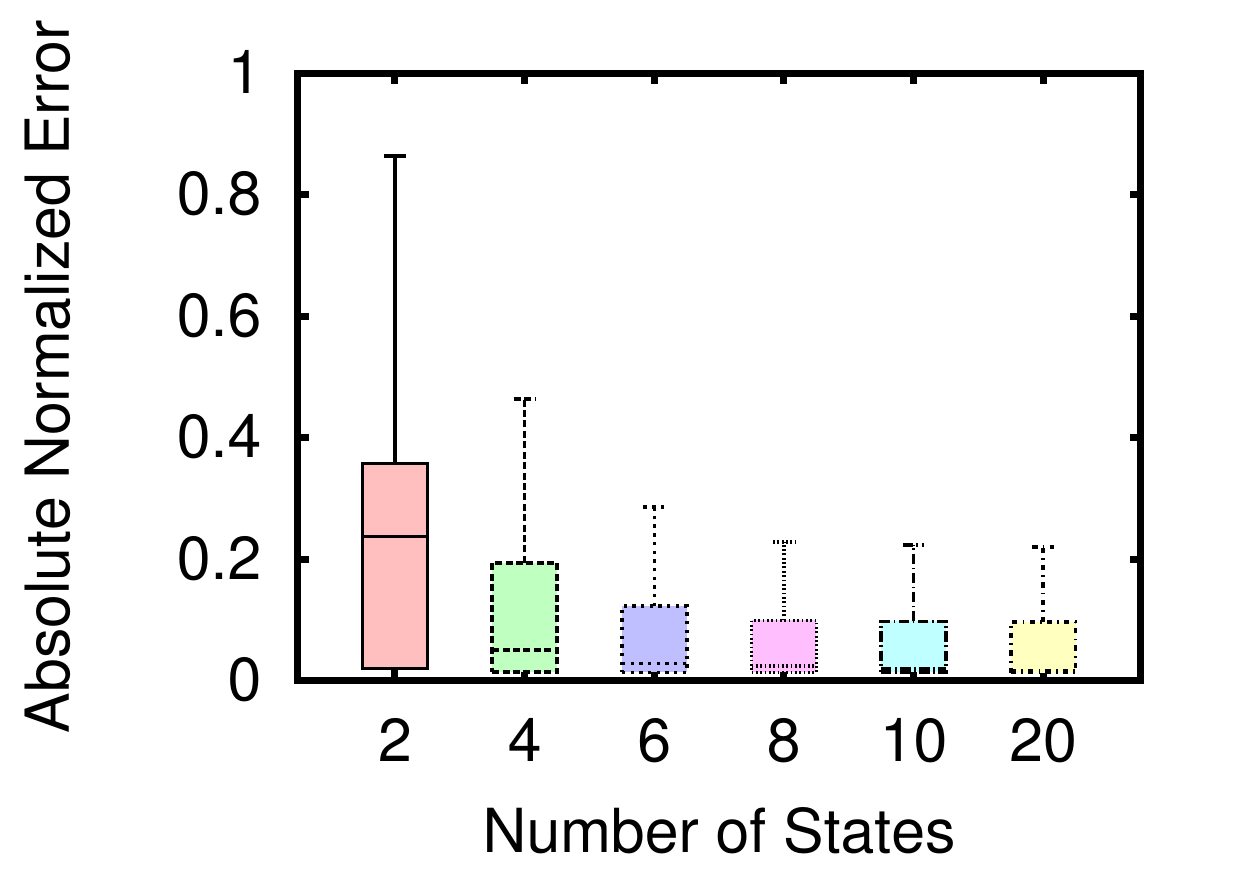}
\tightcaption{Error vs.\ HMM model size.}
\label{fig:hmmstate}
\end{figure}

\begin{figure}[t]
\vspace{-0.5cm}
\centering
\subfloat[90\%ile per session, and median across sessions]
{
\includegraphics[width=0.5\linewidth]{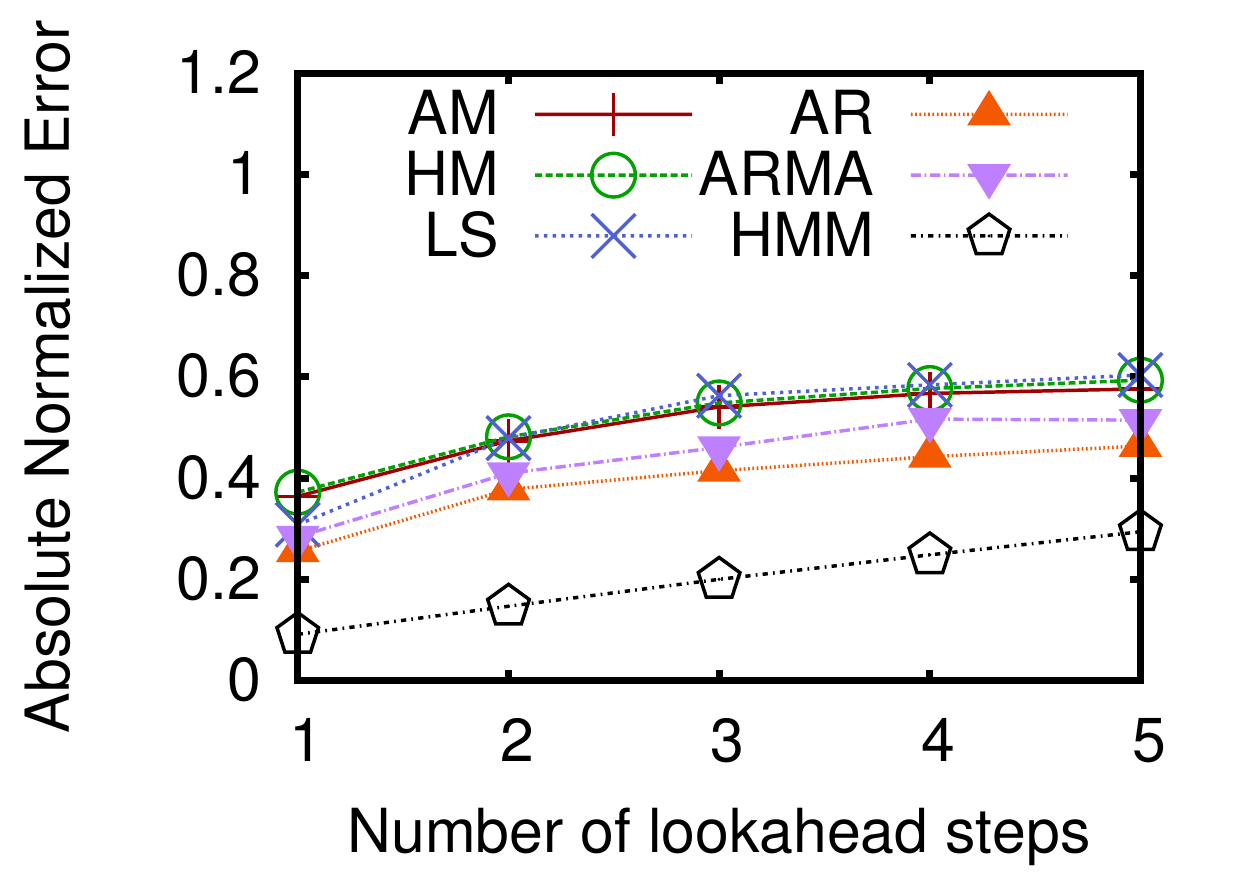}
\label{fig:error:9050}
}
\subfloat[Error distribution (25, 50, 75\%ile)]
{
\includegraphics[width=0.5\linewidth]{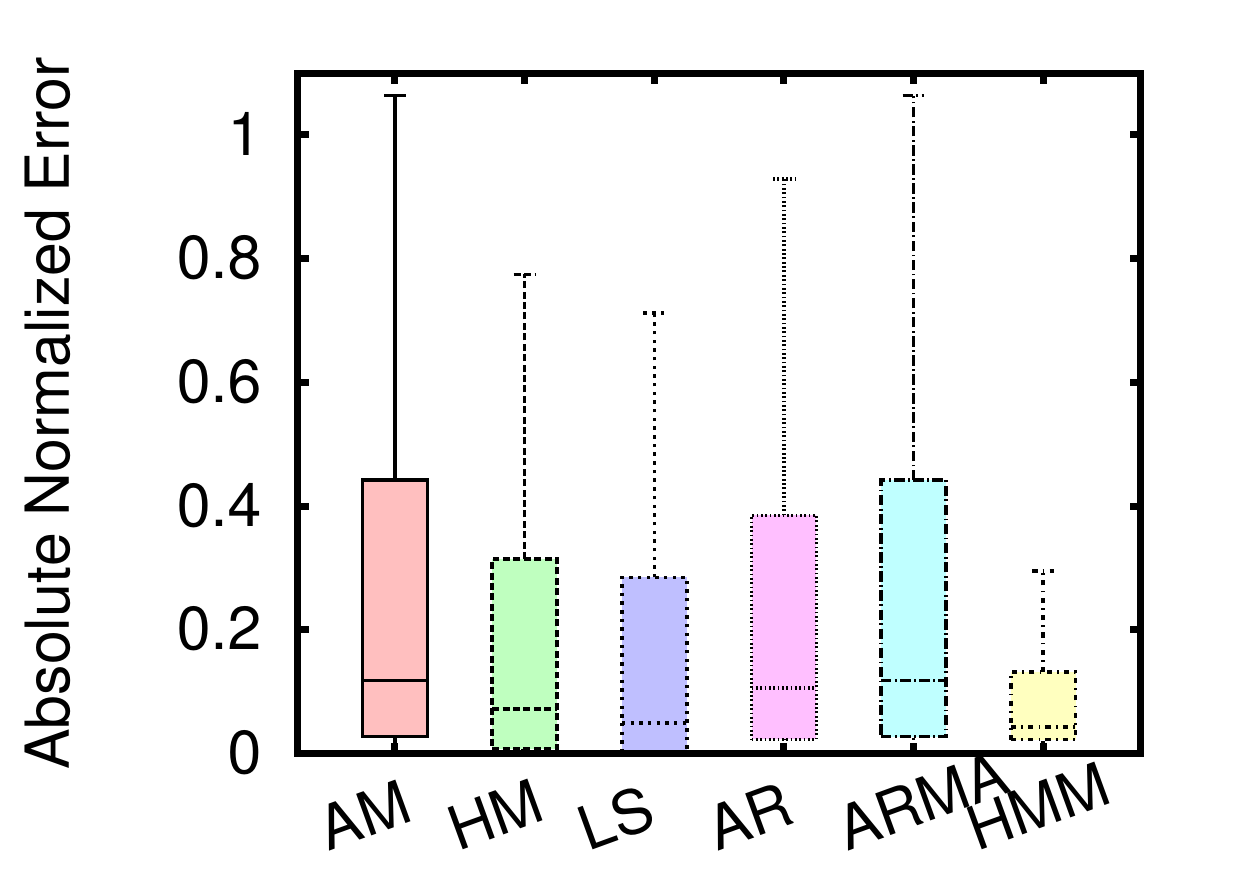}
\label{fig:error:all}
}
\tightcaption{Prediction accuracy of different models; LS=Last Sample, AM=ArithmeticMean, HM=Harmonic Mean, AR =AutoRegressive, ARMA = AutoRegressive Moving Average.}
\vspace{-0.2cm}
\label{fig:error}
\end{figure}

\mypara{Setup} To learn the various parameters (e.g., $\theta_{ARMA}$,
$\theta_{HMM}$), we divide the dataset into equally-sized training and testing
datasets.  We learn these parameters  from training dataset and report error
metrics on  the testing dataset. 
For AR model, we empirically tried different $p$ values in the training dataset and found
$p=5$ yields the best result.
\vspace{-0.2cm}

\mypara{Error metric} For each slot $t$ of a  session $s$, we compute the
absolute normalized error
$\mathit{Err}_{s,t}=\frac{|\hat{W}_{s,t}-W_{s,t}|}{W_{s,t}}$, where
$\hat{W}_{s,t}$ and $W_{s,t}$ denote the predicted and true throughput for slot
$t$ of session $s$. Given these ``atomic'' error values, we can summarize the
error within and across sessions in different ways; e.g., median per-session
and median across sessions or median per-session and 90-th percentile across
sessions.
\vspace{-0.2cm}

\mypara{Configuring HMM} One natural question about the HMM is how many states $M$
we need in practice.  Having more complex models with more states can decrease
the error, but also increases the training time and risks of overfitting.
  Figure~\ref{fig:hmmstate} shows the testing error for  HMMs with varying number of
states.  We see that while the
error decreases with more complex state models, we see a natural diminishing
returns property after 6 states.  As a practical tradeoff between the above
considerations, we choose a 6-state HMM.
\vspace{-0.2cm}

\mypara{HMM vs.\ Strawman solutions}  Figure~\ref{fig:error} considers two
possible ways to summarize the per-slot error values.
Figure~\ref{fig:error:9050} shows the median across sessions of the ``tail''
90-percentile prediction error within  a session and Figure~\ref{fig:error:all}
shows the  overall distribution of the per-slot error values. In both cases, we
see that the HMM model clearly outperforms other techniques. For instance,  in
Figure~\ref{fig:error:9050}, the HMM approach has 60\% improvement over the
second best predictor (AR). Similarly, in the overall distribution we see that
the HMM dramatically reduces the tail of the errors; e.g., more than 75\% of
the predictions of HMM have less than $<$18\% compared to $\ge$27\% for other
models. We also considered other summarizations such as median across sessions
of per-session median,   average-of-average etc., and found consistent results
that HMM significantly outperforms the strawman models (not shown). Note that
 the expected benefits of HMM predictors will be even bigger when we go to
finer time-scale, e.g. second-level instead of minute-level throughput prediction.

One minor downside is that the ``low tail'' (25 percentile) error of HMM is
worse than the last-sample predictor.  This is due to some highly stable
sessions where throughput is  constant and thus last-sample predictor has zero
error.  Due to the quantization effect with only 6 states, there is a small
bias with HMM predictions. However,  as we will see next this has no impact on
the application quality of experience.

\subsection{ Improvement in video QoE}

Next, we evaluate the improvement in user quality of experience (QoE) gained by
using the improved HMM-based throughput prediction in the context of dynamic
adaptive streaming over HTTP (DASH)~\cite{yin2015controlvideo, jiang2014improving}.
\vspace{-0.15cm}

\mypara{Setup} Our goal is to evaluate the benefit of improved
throughput prediction via HMM and not to evaluate the specific video adaptation
heuristics or artifacts. To this end,  for the adaptation algorithm we follow
 strategies  formulated  by recent efforts~\cite{yin2015controlvideo,hotmobile}, that
  take as input  throughput predictions for the next few epochs (e.g., via
harmonic mean) and solve an exact integer linear programming optimization to
decide the bitrate for the next chunk. As a baseline, we also consider the
buffer-based (BB) policy which does not use any throughput
prediction~\cite{huang2014dash}.
\vspace{-0.15cm}

\mypara{Error metric} Identifying suitable QoE functions for video is an open problem~\cite{videoqoehotnets}.
Here,  we adopt a simple linear  model suggested by previous
work~\cite{yin2015controlvideo}, which is the weighted sum of different
factors  such as average video quality, average quality variation, and total
rebuffer time.  We compute  a {\em  normalized QoE} metric of each algorithm
relative  to the theoretical optimal, which could be achieved with the perfect
knowledge of future throughput.
\vspace{-0.15cm}

\begin{figure}[t]
\vspace{-0.5cm}
\centering
\includegraphics[width=0.8\linewidth]{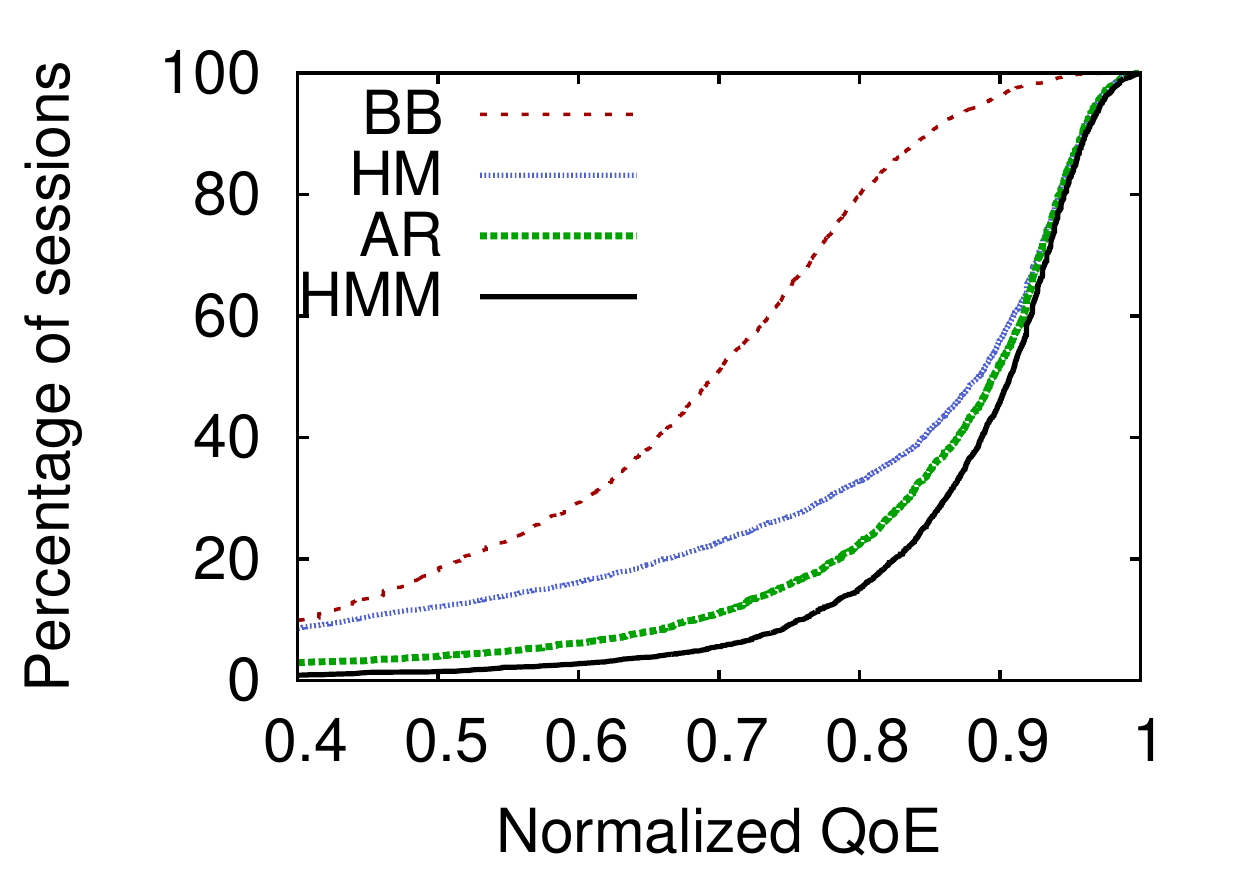}
\vspace{-0.1cm}
\tightcaption{Adaptive video QoE improvement using throughput prediction; BB refers to the pure buffer-based adaptation  that ignores throughput~\cite{huang2014dash}.}
\label{fig:qoe:normalized}
\end{figure}

\mypara{QoE improvement} Figure~\ref{fig:qoe:normalized} shows the CDF of the
normalized QoE of different approaches. For clarity, we focus
on a subset of predictors since the lines of other strawman solutions are very
close to the Harmonic mean (HM) and AR. First, the result confirms observations from
prior work that accurate prediction can dramatically improve QoE over the
baseline buffer-based approach~\cite{yin2015controlvideo,huang2014dash}.
Second, we also see the improved prediction accuracy of HMM also  leads to the
best QoE especially in the lower tail; e.g., the gap between the 20\%ile QoE of
HMM and the harmonic mean suggested in prior work~\cite{jiang2014improving} is
almost 25\%.\footnote{One subtle issue is that even though AR is worse than Harmonic Mean
in terms of prediction error its QoE distribution is better. This is due to a
combination of two factors. First, the AR algorithm tends to be conservative
and underestimates throughput; thus its rebuffering is low. Second, in our
normalized QoE rebuffering has a relatively higher weight. Together, AR's QoE
is better.}    Third, we see that the HMM-based approach is also very close to
the optimal QoE achievable with perfect knowledge, with median being 90\% of
the optimal.
\vspace{-0.05cm}

\section{Conclusions}
\label{sec:conclusion}

 Our imminent need for understanding  throughput stability and predictability
 is motivated by adaptive streaming over
HTTP~\cite{yin2015controlvideo,tian2012towards,hotmobile}.  There is surprisingly little
work on this topic and large-scale datasets on ``long lived''
sessions with continuous throughput measurements needed to shed
 light on these aspects  appear to be especially scarce.\footnote{Perhaps the paucity stems from the fact that
throughput stability is not necessary for previous ``killer apps'' such
as Web or file transfer.} Our work  bridges this gap by (1) providing a
large-scale measurement analysis of intra-session throughput stability and (2)
an online prediction mechanism based on a 
hidden Markov model.  We hope
that our work inspires further research on this topic  at more fine-grained
timescales and across different deployment
scenarios (e.g., cellular).

\newcommand\blfootnote[1]{%
  \begingroup
  \renewcommand\thefootnote{}\footnote{#1}%
  \addtocounter{footnote}{-1}%
  \endgroup
}

{
\scriptsize
\bibliographystyle{abbrv}
\bibliography{conext2015new}

\begin{thebibliography}{10}

\bibitem{ciscovni}
{Cisco Visual Networking Index}.
\newblock \url{
  http://www.cisco.com/c/en/us/solutions/service-provider/visual-networking-index-vni/index.html}.

\bibitem{fccmba}
{FCC Measuring Broadband America }.
\newblock \url{http://www.fcc.gov/measuring-broadband-america}.

\bibitem{hsdpa}
{HSDPA}.
\newblock \url{http://home.ifi.uio.no/paalh/dataset/hsdpa-tcp-logs/}.

\bibitem{pathchar}
Pathchar.
\newblock \url{ http://www.caida.org/tools/utilities/others/pathchar/}.

\bibitem{pptv}
{PPTV}.
\newblock \url{http://www.pptv.com/}.

\bibitem{videoqoehotnets}
A.~Balachandran et~al.
\newblock {A Quest for an Internet Video Quality-of-Experience Metric}.
\newblock In {\em HotNets}, 2012.

\bibitem{wideareastability}
H.~Balakrishnan et~al.
\newblock {Analyzing Stability in WideArea Network Performance}.
\newblock In {\em Proc.\ ACM SIGMETRICS}, 1997.

\bibitem{bishop2006pattern}
C.~M. Bishop.
\newblock {\em Pattern recognition and machine learning}.
\newblock springer, 2006.

\bibitem{dischinger2010glasnost}
M.~Dischinger et~al.
\newblock Glasnost: enabling end users to detect traffic differentiation.
\newblock In {\em Proc.\ NSDI}, 2010.

\bibitem{dobrian2011understanding}
F.~Dobrian et~al.
\newblock Understanding the impact of video quality on user engagement.
\newblock In {\em Proc.\ SIGCOMM}, 2011.

\bibitem{tcppredictability}
Q.~He et~al.
\newblock {On the predictability of large transfer TCP throughput}.
\newblock In {\em Proc.\ ACM SIGCOMM}, 2005.

\bibitem{hu2004pathneck}
N.~Hu et~al.
\newblock Locating internet bottlenecks: Algorithms, measurements, and
  implications.
\newblock In {\em Proc.\ SIGCOMM}, 2004.

\bibitem{huang2014dash}
T.~Huang et~al.
\newblock {A Buffer-Based Approach to Rate Adaptation: Evidence from a Large
  Video Streaming Service}.
\newblock In {\em Proc.\ SIGCOMM}, 2014.

\bibitem{confused}
T.-Y. Huang et~al.
\newblock {Confused, Timid, and Unstable: Picking a Video Streaming Rate is
  Hard}.
\newblock In {\em Proc.\ IMC}, 2012.

\bibitem{jain2003end}
M.~Jain et~al.
\newblock End-to-end available bandwidth: measurement methodology, dynamics,
  and relation with tcp throughput.
\newblock {\em IEEE/ACM Transactions on Networking}, 11(4):537 -- 549, 2003.

\bibitem{jiang2014improving}
J.~Jiang et~al.
\newblock {Improving Fairness, Efficiency, and Stability in HTTP-Based Adaptive
  Video Streaming with Festive}.
\newblock {\em IEEE/ACM Transactions on Networking}, 22(1):326 -- 340, 2014.

\bibitem{mirza2007machine}
M.~Mirza et~al.
\newblock A machine learning approach to tcp throughput prediction.
\newblock In {\em Proc.\ SIGMETRICS}, 2007.

\bibitem{padhye1998modeling}
J.~Padhye et~al.
\newblock {Modeling TCP throughput: A Simple Model and its Empirical
  Validation}.
\newblock In {\em Proc.\ SIGCOMM}, 1998.

\bibitem{paxson1996sigcomm}
V.~Paxson.
\newblock {End-to-End Routing Behavior in the {I}nternet}.
\newblock Aug. 1996.

\bibitem{rexford2002bgp}
J.~Rexford et~al.
\newblock {BGP} routing stability of popular destinations.
\newblock In {\em Proc.\ SIGCOMM IMC}, 2002.

\bibitem{bismark}
S.~Sundaresan et~al.
\newblock {Broadband Internet Performance: A View From the Gateway}.
\newblock In {\em Proc.\ SIGCOMM}, 2011.

\bibitem{tian2012towards}
G.~Tian et~al.
\newblock Towards agile and smooth video adaptation in dynamic http streaming.
\newblock In {\em Proc.\ CoNEXT}, 2012.

\bibitem{yin2015controlvideo}
X.~Yin et~al.
\newblock {Toward a Principled Framework to Design Dynamic Adaptive Streaming
  Algorithms over HTTP}.
\newblock In {\em Proc.\ HotNets}, 2014.

\bibitem{zhang2001constancy}
Y.~Zhang et~al.
\newblock On the constancy of {I}nternet path properties.
\newblock In {\em Proc.\ IMW}, 2001.

\bibitem{hotmobile}
X.~K. Zou et~al.
\newblock {Can Accurate Predictions Improve Video Streaming in Cellular
  Networks?}
\newblock In {\em HotMobile}, 2015.

\end{thebibliography}
}



\end{document}